\pdfoutput=1
\documentclass[11pt]{article}

\usepackage[letterpaper,margin=1in]{geometry}
\usepackage[T1]{fontenc}
\usepackage[utf8]{inputenc}
\usepackage{lmodern}
\usepackage{amsmath,amsfonts,amssymb}
\usepackage{graphicx}
\usepackage{booktabs}
\usepackage{multirow}
\usepackage{url}
\usepackage{cite}
\usepackage{authblk}
\usepackage[colorlinks=true, allcolors=blue]{hyperref}

\date{}
\newcommand{\keywords}[1]{\par\vspace{0.6em}\noindent\textbf{Keywords: }#1\par}
\newcommand{\acknowledgments}{\section*{Acknowledgments}}

\title{Digital twins for compact hybrid quantum--classical learning in FMCW radar detection}
\author[1,*]{Sebastian Ratto V.}
\author[1]{Ahmed N.\ Sayed}
\author[2]{Arien Sligar}
\author[3]{Jose R.\ Rosas--Bustos}
\author[1]{Omar M.\ Ramahi}
\author[1]{George Shaker}

\affil[1]{Department of Electrical and Computer Engineering, University of Waterloo, 200 University Avenue West, Waterloo, ON N2L 3G1, Canada}
\affil[2]{Synopsys, Inc., Mountain View, CA, United States}
\affil[3]{Applied Quantum Technologies Institute (AQTI), Waterloo, ON, Canada}
\affil[*]{Corresponding author: srattova@uwaterloo.ca}

\begin{document}
\maketitle

\begin{abstract}
Frequency-modulated continuous-wave radar sensing often relies on labeled measurements that are costly, restricted, or difficult to collect at scale. This work evaluates physics-informed digital twins as controlled testbeds for early-stage quantum--classical radar learning. Two synthetic radar benchmarks are considered: unmanned aerial vehicle classification from range--Doppler maps and human fall detection from Doppler--time spectrograms. For both tasks, inputs are standardized, reduced using principal component analysis, and classified using either a radial basis function support vector classifier or a quantum support vector classifier. All quantum-kernel results are obtained using noiseless classical simulation; no quantum hardware is used, and no quantum-advantage claim is made. Across five random seeds, the quantum support vector classifier improves the UAV benchmark from four principal components onward, reaching an accuracy of $0.941 \pm 0.012$ at eight components, compared with $0.880 \pm 0.029$ for the classical baseline. On the fall-detection benchmark, both classifiers perform similarly, with a small quantum-kernel improvement at higher feature dimensions. A Gaussian-noise robustness study shows limited performance degradation across the tested noise levels, while preserving the UAV quantum-kernel gain. These results support digital twins as useful, controlled environments for radar-QML benchmarking prior to measured-data validation and hardware execution.
\end{abstract}


\section{INTRODUCTION}

Frequency-modulated continuous-wave (FMCW) radar provides contactless sensing for privacy-preserving healthcare monitoring, autonomous perception, and defence-relevant detection of small unmanned aerial vehicles (UAVs).\cite{clemente2021,fioranelli2020} In each of these domains, labelled radar measurements remain a practical bottleneck. Clinically relevant falls are difficult to record ethically and at scale,\cite{who_falls_2021,cdc_falls_facts_2024,Elbadrawy2023MapCon,ratto2026radar} while UAV signatures depend on airframe geometry, rotor motion, radar configuration, trajectory, and operating environment.\cite{caris2017,Sayed2024MicrowaveMagUAVDigitalTwins,Sayed2024SensorsLettersISARDigitalTwins} These constraints motivate the use of physics-informed digital twins (DTs), which generate radar returns from explicit scene geometry and electromagnetic modelling rather than from purely statistical augmentation.

DT-based radar studies have been reported for UAV classification, inverse synthetic aperture radar imaging, MIMO sensing, and noise-aware simulation.\cite{Sayed2024MicrowaveMagUAVDigitalTwins,Sayed2024DronesReview,Sayed2024SensorsLettersISARDigitalTwins,Sayed2024TMTTMIMOUAV,Ahmed2025NoiseUAVDigitalTwins} Related DT workflows have also been used in healthcare and indoor sensing, including fall monitoring, human activity representation, and people-counting applications.\cite{Elbadrawy2023MapCon,ratto2025enhancing,ratto2025digital_xr,ratto2025digital_antem,ratto2026radar,ratto2025radar} In this paper, we use DT-generated radar products not as a substitute for deployment validation, but as controlled benchmark inputs. This distinction is important: synthetic data let us isolate modelling choices under matched conditions, but measured data are still required before deployment-oriented conclusions can be drawn.

Quantum machine learning (QML) offers feature-map and kernel constructions that can be evaluated under small-dimensional input bottlenecks.\cite{havlicek2019,schuld2019quantum} Radar is a relevant domain for this type of benchmark because FMCW sensing begins with complex-baseband measurements containing amplitude, phase, Doppler evolution, and interference structure. A quantum feature map also represents data through rotations and phase-dependent transformations before comparing encoded states through an inner product. However, the present study uses magnitude range--Doppler maps and Doppler--time spectrograms rather than raw complex I/Q sequences. Therefore, the comparison does not test whether quantum circuits preserve or exploit radar phase information directly. It tests whether a rotation-based quantum kernel separates the same compact radar features differently from a conventional Euclidean RBF kernel.

The central benchmarking question is practical: under identical preprocessing, identical train/test splits, and identical feature dimensions, does a fidelity-based quantum kernel improve classification relative to a strong classical kernel? This question is difficult to answer when studies change the dataset, feature dimension, preprocessing, classifier, and evaluation protocol simultaneously. We therefore use a matched pipeline in which the radar representation differs by task, but the learning protocol after radar-product formation is the same.

This paper presents a controlled comparison of a classical RBF kernel and a quantum fidelity kernel on two DT-generated FMCW radar tasks. The first task classifies three UAV airframes from range--Doppler maps. The second task classifies fall and non-fall examples from Doppler--time spectrograms. For both tasks, we flatten the radar product, standardize the input using training-set statistics, project the data to a low-dimensional PCA bottleneck, and train either RBF--SVC or QSVC. We evaluate all quantum kernels by noiseless statevector simulation in Qiskit; no quantum hardware is used.

The contributions are threefold. First, we organize two DT-generated radar tracks into a single compact kernel-benchmarking pipeline for UAV classification and fall detection. Second, we report repeated-seed RBF--SVC and QSVC comparisons across PCA dimensions $d \in \{2,4,6,8\}$, showing that QSVC improves the UAV benchmark only after the bottleneck retains sufficient class structure and that the effect is task-dependent. Third, we evaluate raw-input Gaussian perturbations at test time, showing that both kernels remain stable over the tested noise sweep and that the UAV QSVC advantage is preserved. We frame these results as quantum-utility evidence for a controlled DT benchmark, not as evidence of quantum advantage, hardware speedup, or deployment readiness.

\section{METHODOLOGY}

\subsection{FMCW radar products}

An FMCW radar transmits linear chirps whose frequency changes over a chirp duration $T_c$. For bandwidth $B$, the sweep slope is $S=B/T_c$. After dechirping, the beat frequency contains range information, and the phase progression from chirp to chirp contains Doppler information. In the simplified narrowband interpretation used here,
\begin{equation}
S=\frac{B}{T_c},
\qquad
R \approx \frac{c f_b}{2S},
\qquad
v \approx \frac{\lambda f_D}{2},
\label{eq:fmcw_basic}
\end{equation}
where $c$ is the speed of light, $\lambda$ is the radar wavelength, $f_b$ is the beat frequency, and $f_D$ is the Doppler frequency. These quantities connect the received baseband signal to physically meaningful range and velocity axes.\cite{clemente2021,fioranelli2020}

We use different radar products for the two tasks because the underlying physical signatures differ. For UAV classification, each input is a range--Doppler map (RDM). The RDM captures the distribution of reflected energy over range and radial velocity within a coherent processing interval, which is appropriate for airframes whose body and rotor components generate structured Doppler signatures. We form an RDM by applying a fast-time Fourier transform for range and a slow-time Fourier transform across chirps for Doppler:
\begin{equation}
\mathrm{RDM}(k,\ell)
=
\left|
\mathrm{FFT}_{\mathrm{slow}}
\left\{
\mathrm{FFT}_{\mathrm{fast}}\{x[n,m]\}
\right\}
\right|,
\label{eq:rdm}
\end{equation}
where $k$ indexes range bin and $\ell$ indexes Doppler bin.

For fall detection, a single range--Doppler snapshot is less informative because the event is defined by motion evolution over time. We therefore represent each fall input as a Doppler--time spectrogram formed by stacking Doppler profiles over consecutive short windows:
\begin{equation}
\mathrm{Spec}(\ell,t)
=
\left|
\mathrm{FFT}_{\mathrm{slow}}\{x_t[m]\}
\right|,
\label{eq:spectrogram}
\end{equation}
where $\ell$ indexes Doppler bin and $t$ indexes the time frame. The UAV input therefore emphasizes range and velocity structure over one coherent interval, while the fall input emphasizes velocity evolution over a motion window. After product formation, both inputs are two-dimensional arrays and can be processed by the same learning pipeline.

\subsection{Radar configurations}
\label{sec:radar_config}

The two DT tracks use different FMCW radar configurations, summarized in Table~\ref{tab:radar_config}. We do not force the front ends to be identical because the two sensing problems require different radar-product definitions and operating regimes. Instead, we match the learning benchmark after product formation by applying the same standardization, PCA bottleneck, and classifier protocol to both tracks. This design isolates the comparison between kernels while respecting the physical differences between UAV RDMs and fall spectrograms.

\begin{table}[t]
\centering
\caption{Radar parameters and input products used in the two digital-twin tracks.}
\label{tab:radar_config}
\small
\renewcommand{\arraystretch}{1.12}
\setlength{\tabcolsep}{6pt}
\begin{tabular}{@{}lcc@{}}
\toprule
Parameter & UAV track & Fall track \\
\midrule
Carrier frequency & $77~\mathrm{GHz}$ & $60~\mathrm{GHz}$ \\
Bandwidth & $300~\mathrm{MHz}$ & $499.7~\mathrm{MHz}$ \\
Range resolution & $\approx 0.5~\mathrm{m}$ & $\approx 0.30~\mathrm{m}$ \\
Velocity resolution & $\approx 0.4~\mathrm{m/s}$ & $\approx 0.095~\mathrm{m/s}$ \\
Radar product & Range--Doppler map & Doppler--time spectrogram \\
Input size & $128 \times 510$ & $256 \times 64$ \\
\bottomrule
\end{tabular}
\end{table}

\subsection{UAV digital-twin dataset}

The UAV dataset contains 600 labelled RDMs evenly distributed across three classes: BlackEagle helicopter, DJI~S900 hexacopter, and Phantom~3 quadcopter, with 200 samples per class. The RDMs are generated by a high-frequency electromagnetic DT that combines shooting-and-bouncing-rays propagation with detailed CAD geometry for each airframe.\cite{Sayed2024MicrowaveMagUAVDigitalTwins,Sayed2024SensorsLettersISARDigitalTwins,Sayed2024DronesReview} Prior UAV radar-DT work has shown that geometry, flight regime, radar configuration, MIMO processing, and noise level can affect classification performance.\cite{Sayed2024MicrowaveMagUAVDigitalTwins,Sayed2024SensorsLettersISARDigitalTwins,Sayed2024TMTTMIMOUAV,Ahmed2025NoiseUAVDigitalTwins} We reuse the generated RDMs without changing the radar-scene generation step, so the present study focuses on the compact classifier head rather than on DT synthesis itself.

We choose this dataset because it provides a controlled three-class benchmark with physically distinct airframes and matched sample counts. An alternative would be to train directly on measured UAV flights, but measured data introduce uncontrolled variation in trajectory, clutter, range, aspect angle, and environmental conditions. Those factors are essential for deployment validation, but they would make it harder to isolate the effect of the kernel under a compact PCA bottleneck. The DT setting therefore supports an early-stage comparison in which the input generation process is fixed and the classifier branch is the primary variable.

\subsection{Fall digital-twin dataset}

The fall dataset is generated using a $60~\mathrm{GHz}$ FMCW DT of an assisted-living-style indoor environment, following prior radar-DT workflows for healthcare monitoring and fall representation studies.\cite{Elbadrawy2023MapCon,ratto2025enhancing,ratto2025digital_xr,ratto2025digital_antem,ratto2026radar} The source data contain 16 simulated radar clips covering fall and non-fall activities. Each clip is stored as a four-dimensional cube $\texttt{rdm\_db}(T, \mathrm{RX}, D, R)$ over slow-time frames, receive channels, Doppler bins, and range bins. We average across receive channels before temporal stacking, then form per-window Doppler--time spectrogram inputs by stacking Doppler magnitude profiles across $T=64$ consecutive slow-time frames. The resulting export contains 286 fall examples and 269 no-fall examples, for 555 samples in total.

We use clip-level partitioning for all fall train/test splits. This choice prevents spectrogram windows derived from the same source clip from appearing in both training and testing, which would otherwise create temporal leakage. A random window-level split would produce more optimistic performance estimates because adjacent windows from the same simulated motion can share trajectory structure. The clip-level split is therefore stricter and better aligned with the intended evaluation question: whether the compact classifier generalizes across held-out simulated clips rather than memorizing local window variants.

\subsection{Matched preprocessing and compact feature bottleneck}
\label{sec:pipeline}

\begin{figure}[t]
\centering
\includegraphics[width=\linewidth]{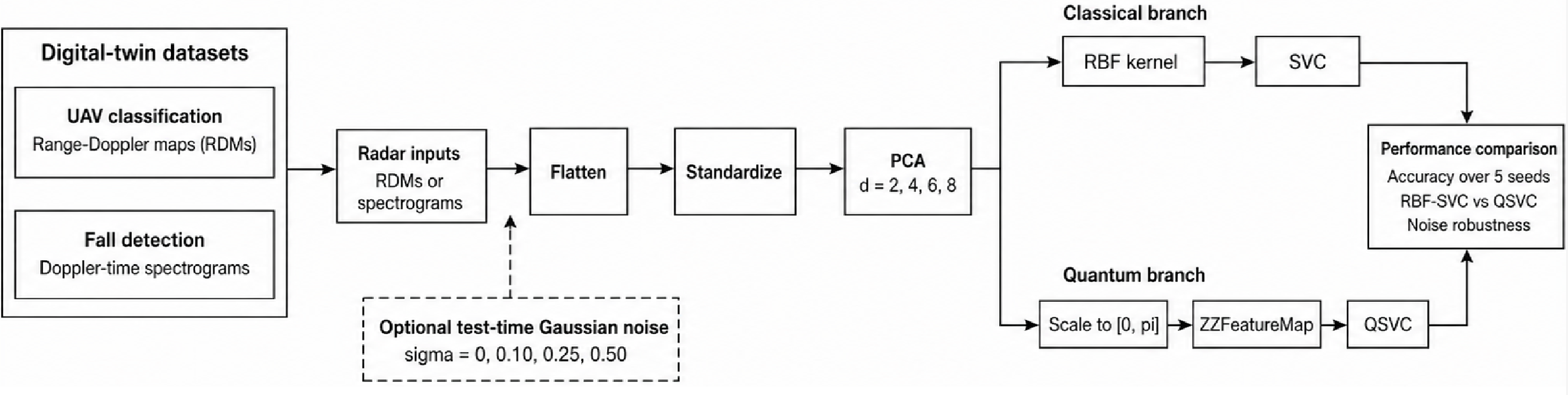}
\caption{Matched compact kernel pipeline. Each radar input is flattened, standardized using training-set statistics, projected by PCA, and classified with either a classical RBF kernel or a fidelity-based quantum kernel.}
\label{fig:pipeline}
\end{figure}

Both datasets are processed by the same pipeline, summarized in Fig.~\ref{fig:pipeline}. Each input $\mathbf{X}$ is flattened to a vector $\mathbf{x} \in \mathbb{R}^{D}$. For the UAV track, $D=128 \times 510=65{,}280$; for the fall track, $D=256 \times 64=16{,}384$. We standardize features using training-set statistics only and project them by principal component analysis (PCA)\cite{pearson1901,sklearn} to $d \in \{2,4,6,8\}$.

We use PCA for two reasons. First, quantum kernel simulation and near-term quantum hardware both impose strong dimensional constraints, since each retained feature maps to a qubit in the selected feature map. Second, PCA provides a deterministic, unsupervised bottleneck that can be applied identically to both kernels, reducing the risk that the comparison is driven by different feature engineering rather than by the kernel geometry. We considered using supervised feature selection or a learned neural embedding, but those alternatives would introduce additional trainable components and make the kernel comparison less direct.

\subsection{Classical RBF--SVC branch}

The classical baseline is an RBF--SVC trained on the PCA features. We select this baseline because the RBF kernel is a strong conventional nonlinear kernel for compact continuous features and is widely used as a default comparator in kernel-method studies. The RBF kernel also provides a meaningful contrast against the fidelity kernel: both define nonlinear similarity functions over the same PCA coordinates, but they impose different geometries. We use the standard \verb|gamma="scale"| heuristic rather than tuning a separate hyperparameter grid, because the goal is a matched compact benchmark rather than a search for the best possible classical model. A larger hyperparameter sweep could improve the RBF branch, but it would also change the comparison from a controlled kernel study into a model-selection study.

\subsection{Quantum fidelity-kernel branch}

The quantum branch uses a QSVC with a \texttt{ZZFeatureMap} fidelity statevector kernel. Before quantum encoding, we rescale the PCA features to $[0,\pi]$ using a min--max transform fitted on the training set. We apply this step because the feature map uses inputs as circuit rotation angles; without a fixed angular range, differences in PCA scale could dominate the encoded state geometry. The same fitted transform is then applied to the corresponding test features.

The \texttt{ZZFeatureMap} with \verb|reps=1| prepares a data-dependent state on $d$ qubits using Hadamard gates, single-qubit phase rotations parameterized by the features, and ZZ-type entangling rotations parameterized by pairwise feature functions.\cite{havlicek2019} If $U_\phi(\mathbf{x})$ denotes the resulting feature-map unitary, the quantum kernel between two samples is the squared fidelity between the encoded states:\cite{havlicek2019,schuld2019quantum}
\begin{equation}
K_Q(\mathbf{x}, \mathbf{y})
=
\left|
\left\langle 0^{\otimes d}\right|
U_\phi^{\dagger}(\mathbf{x})
U_\phi(\mathbf{y})
\left|0^{\otimes d}\right\rangle
\right|^{2}.
\label{eq:kq}
\end{equation}
We compute this kernel exactly with Qiskit's \verb|FidelityStatevectorKernel|.\cite{qiskit} The QSVC then trains a classical support vector machine using the precomputed quantum kernel matrix.

This branch is a quantum-kernel model, not a variational quantum classifier. The circuit contains no trainable quantum parameters, and the learned component is the classical SVC optimization on the kernel matrix. We use the \texttt{ZZFeatureMap} because it is a standard nonlinear quantum feature map for kernel studies and because it can be evaluated at the small feature dimensions considered here. We use \verb|reps=1| to keep the circuit compact and to avoid conflating the kernel comparison with depth-induced simulation or hardware constraints. Deeper circuits may change separability, but they would also increase computational cost and, on hardware, noise sensitivity.

\subsection{Repeated-seed evaluation}

All experiments are repeated over five seeds, $\{7,11,21,42,84\}$, with $25\%$ of samples held out for testing. For the fall task, the holdout is performed at the clip level. For each seed, task, bottleneck dimension, and classifier branch, we record test accuracy. We report the mean and standard deviation across seeds.

We use repeated seeds because a single split can overstate or understate performance, especially for the smaller fall dataset. We use a fixed holdout proportion rather than cross-validation because quantum kernel computation scales with the number of pairwise kernel evaluations, and repeated cross-validation would substantially increase runtime. The five-seed protocol provides a practical compromise: it exposes split sensitivity while keeping the statevector-kernel benchmark computationally manageable.

\subsection{Noise injection protocol}
\label{sec:noise_protocol}

Real radar measurements include receiver noise, gain variation, residual clutter, finite-precision effects, and environment-dependent artefacts. To test input-level robustness in a controlled way, we add Gaussian noise directly to the raw test arrays before standardization and PCA. For each test input,
\begin{equation}
\widetilde{\mathbf{X}}
=
\mathbf{X}
+
\boldsymbol{\eta},
\qquad
\boldsymbol{\eta}
\sim
\mathcal{N}(0,(\sigma s_{\mathrm{train}})^2),
\qquad
\sigma \in \{0.0,0.10,0.25,0.50\},
\label{eq:noise}
\end{equation}
where $s_{\mathrm{train}}$ is the standard deviation of all input pixels in the corresponding training set. We express the perturbation scale as a fraction of $s_{\mathrm{train}}$ so that the noise level is comparable across seeds and across tasks with different dynamic ranges.

We perturb only the test inputs. The training set, standardizer, PCA projection, min--max angular transform, and trained classifier remain unchanged from the clean experiment. This protocol asks whether a trained compact pipeline remains stable when the observed input is corrupted at test time. We considered augmenting the training set with noise, but that would test noise-aware training rather than the intrinsic robustness of the already trained pipeline. The Gaussian model is intentionally simple and should not be interpreted as a full radar measurement-noise model.

\section{RESULTS}
\label{sec:results}

\subsection{Representative radar inputs}

\begin{figure}[t]
\centering
\includegraphics[width=0.85\linewidth]{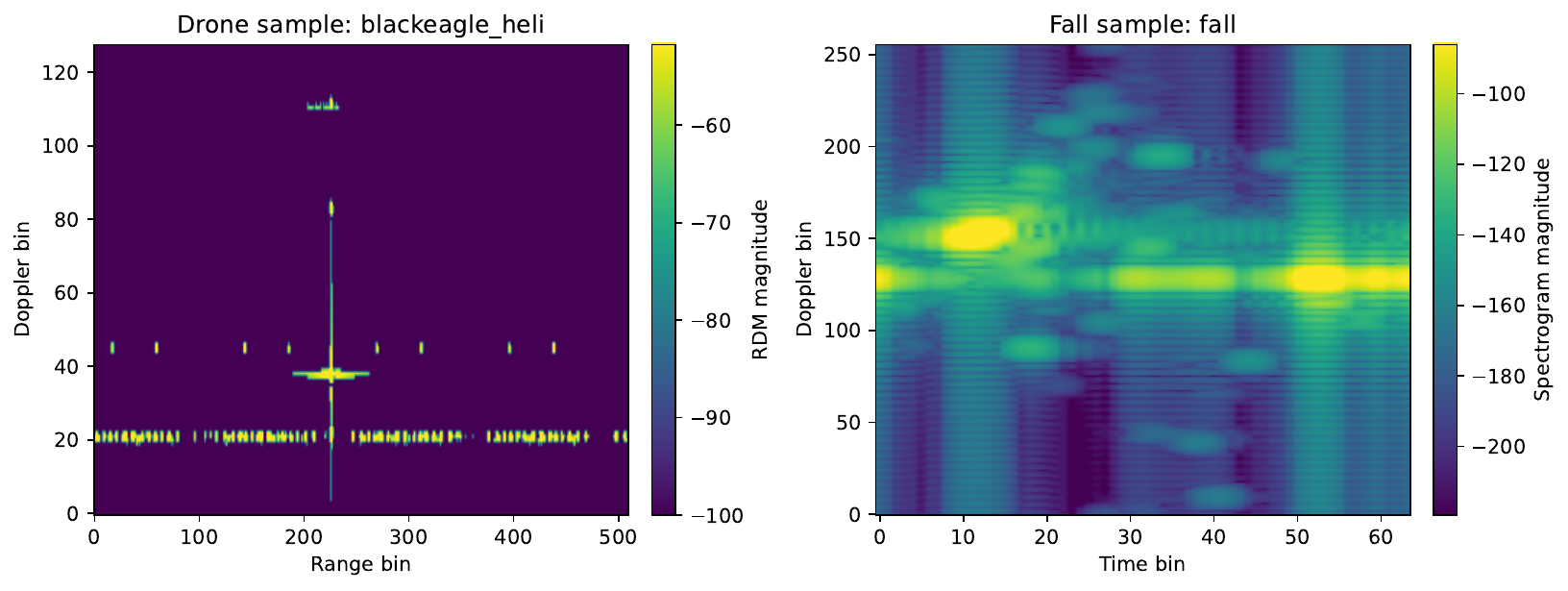}
\caption{Representative radar inputs for the two digital-twin tracks: a UAV range--Doppler map and a fall Doppler--time spectrogram.}
\label{fig:rdm_samples}
\end{figure}

Figure~\ref{fig:rdm_samples} shows representative inputs from the two tracks. The UAV example is an RDM with structured energy across range and Doppler, while the fall example is a Doppler--time spectrogram showing the temporal evolution of body motion. These examples illustrate why the tasks use different radar products before entering the shared learning pipeline.

\subsection{Clean kernel benchmarks}

Table~\ref{tab:results} reports the clean test accuracy across PCA dimensions. On the UAV task, RBF--SVC is stronger at $d=2$, with $0.743 \pm 0.021$ accuracy compared with $0.636 \pm 0.062$ for QSVC. From $d=4$ onward, QSVC has the higher mean accuracy. At $d=8$, QSVC reaches $0.941 \pm 0.012$, compared with $0.880 \pm 0.029$ for RBF--SVC.

On the fall task, the two kernels remain close across all dimensions. RBF--SVC has the higher mean at $d=2$ and $d=4$, while QSVC has the higher mean at $d=6$ and $d=8$. The largest fall-task QSVC mean is $0.836 \pm 0.012$ at $d=8$, compared with $0.823 \pm 0.021$ for RBF--SVC. These differences are smaller than the UAV differences and should be interpreted as task-dependent kernel behaviour rather than as a uniform advantage for the quantum kernel.

\begin{table}[t]
\centering
\caption{Test accuracy (mean $\pm$ standard deviation over five seeds) for RBF--SVC and QSVC across PCA bottleneck dimensions $d$. The better mean within each row is shown in bold.}
\label{tab:results}
\renewcommand{\arraystretch}{1.18}
\setlength{\tabcolsep}{8pt}
\small
\begin{tabular}{@{}c c c c c@{}}
\toprule
\multirow{2}{*}{$d$} & \multicolumn{2}{c}{UAV, 3-class} & \multicolumn{2}{c}{Fall, 2-class} \\
\cmidrule(lr){2-3}\cmidrule(lr){4-5}
 & RBF--SVC & QSVC & RBF--SVC & QSVC \\
\midrule
2 & $\mathbf{0.743 \pm 0.021}$ & $0.636 \pm 0.062$ & $\mathbf{0.724 \pm 0.038}$ & $0.712 \pm 0.027$ \\
4 & $0.869 \pm 0.013$ & $\mathbf{0.911 \pm 0.031}$ & $\mathbf{0.814 \pm 0.031}$ & $0.800 \pm 0.051$ \\
6 & $0.884 \pm 0.027$ & $\mathbf{0.929 \pm 0.011}$ & $0.812 \pm 0.028$ & $\mathbf{0.827 \pm 0.016}$ \\
8 & $0.880 \pm 0.029$ & $\mathbf{0.941 \pm 0.012}$ & $0.823 \pm 0.021$ & $\mathbf{0.836 \pm 0.012}$ \\
\bottomrule
\end{tabular}
\end{table}

\begin{figure}[t]
\centering
\includegraphics[width=0.85\linewidth]{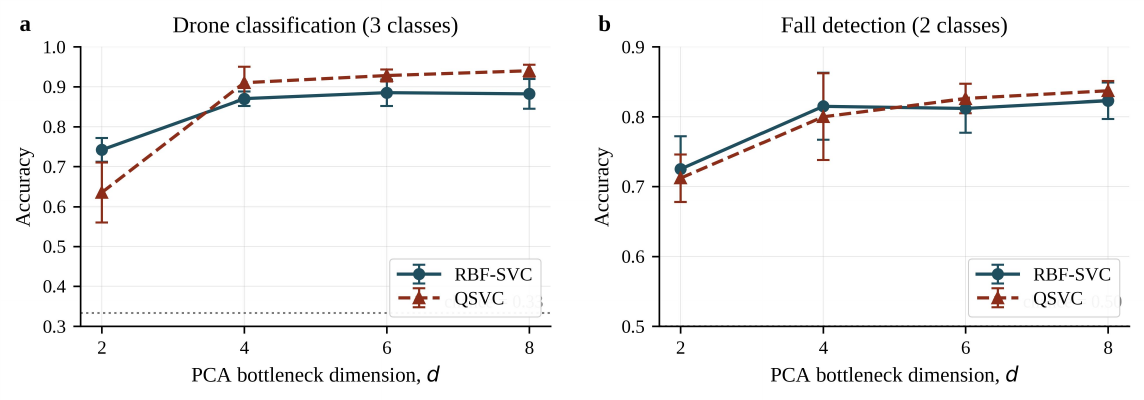}
\caption{Repeated-seed kernel results for the UAV and fall benchmarks across PCA dimension. Markers and bars show the mean and 95\% confidence interval over five seeds.}
\label{fig:metrics}
\end{figure}

Figure~\ref{fig:metrics} visualizes the same trend. The UAV benchmark shows a consistent QSVC gain for $d \geq 4$, including a $+0.061$ mean accuracy gap at $d=8$. In contrast, the fall benchmark shows near-parity between the two kernels, with small QSVC gains only at the larger bottlenecks. The result therefore supports a limited and task-specific quantum-kernel utility claim: the fidelity kernel improves one DT-generated radar benchmark under matched preprocessing, but the benefit does not transfer uniformly to the second radar representation.

\subsection{Noise sensitivity}

The clean benchmarks use noiseless statevector kernel evaluation and do not include measured radar imperfections. We therefore apply the test-time perturbation protocol in Eq.~(\ref{eq:noise}) to both tasks. Gaussian noise is added to the raw test arrays before the clean training-set standardizer, PCA projection, and classifier are applied. This procedure keeps the trained model fixed and evaluates the effect of corrupted test observations.

\begin{table}[t]
\centering
\caption{Noise robustness. Entries are RBF--SVC/QSVC mean accuracy over five seeds; the better value in each pair is shown in bold.}
\label{tab:noise}
\footnotesize
\renewcommand{\arraystretch}{1.08}
\setlength{\tabcolsep}{4pt}
\begin{tabular*}{\textwidth}{@{\extracolsep{\fill}}l c c c c c@{}}
\toprule
Task & $d$ & $\sigma=0.00$ & $\sigma=0.10$ & $\sigma=0.25$ & $\sigma=0.50$ \\
\midrule
\multirow{3}{*}{UAV, 3-class}
& 4 & $0.869/\mathbf{0.911}$ & $0.869/\mathbf{0.908}$ & $0.867/\mathbf{0.913}$ & $0.867/\mathbf{0.904}$ \\
& 6 & $0.884/\mathbf{0.929}$ & $0.883/\mathbf{0.929}$ & $0.881/\mathbf{0.931}$ & $0.879/\mathbf{0.924}$ \\
& 8 & $0.880/\mathbf{0.941}$ & $0.880/\mathbf{0.941}$ & $0.881/\mathbf{0.940}$ & $0.877/\mathbf{0.935}$ \\
\midrule
\multirow{3}{*}{Fall, 2-class}
& 4 & $\mathbf{0.814}/0.800$ & $\mathbf{0.814}/0.803$ & $\mathbf{0.807}/0.800$ & $\mathbf{0.812}/0.794$ \\
& 6 & $0.812/\mathbf{0.827}$ & $0.812/\mathbf{0.827}$ & $0.813/\mathbf{0.825}$ & $0.807/\mathbf{0.827}$ \\
& 8 & $0.823/\mathbf{0.836}$ & $0.825/\mathbf{0.837}$ & $0.825/\mathbf{0.852}$ & $0.830/\mathbf{0.840}$ \\
\bottomrule
\end{tabular*}
\end{table}

Table~\ref{tab:noise} reports mean accuracy versus $\sigma$ for $d \in \{4,6,8\}$. Across both tasks, the changes across noise levels are small. On the UAV task, QSVC remains higher than RBF--SVC at every reported dimension and noise level. On the fall task, the same near-parity pattern from the clean benchmark remains: RBF--SVC is higher at $d=4$, while QSVC is higher at $d=6$ and $d=8$.

The observed robustness should be interpreted within the tested perturbation regime. Because the standardizer and PCA projection are fitted on clean training data, the noisy test inputs are projected into the same compact feature space before either kernel is evaluated. In this experiment, that bottleneck appears to preserve the dominant class structure despite the injected noise. Other perturbations, including coloured noise, clutter ridges, quantization, calibration drift, distribution shift, and phase errors, may affect the two kernels differently.

\subsection{Visualization of noisy inputs}

\begin{figure}[t]
\centering
\includegraphics[width=\linewidth]{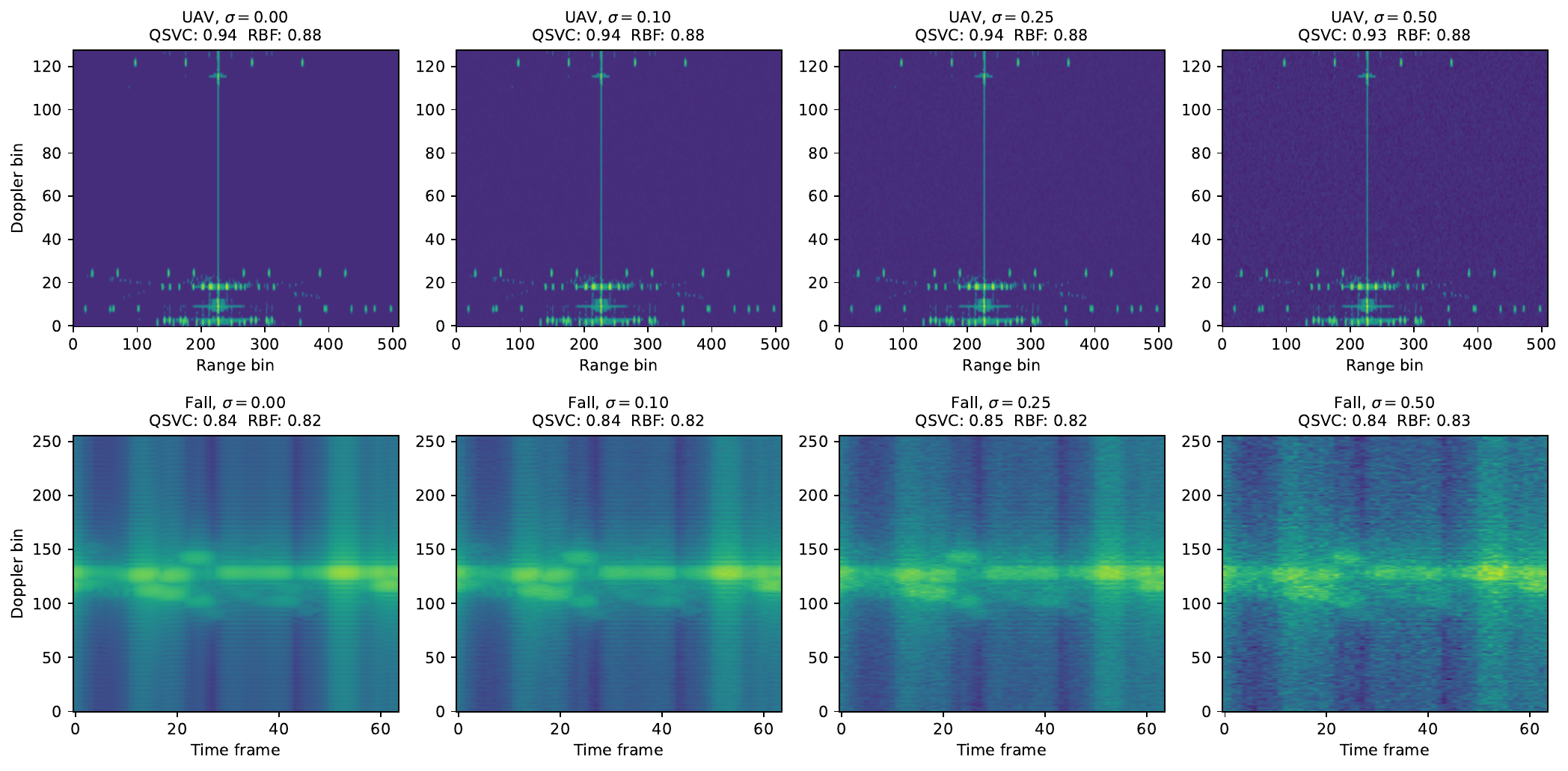}
\caption{Representative UAV RDMs and fall Doppler--time spectrograms under increasing raw-input Gaussian noise.}
\label{fig:noisy_rdms}
\end{figure}

Figure~\ref{fig:noisy_rdms} illustrates the raw-input noise injection for representative UAV RDMs and fall Doppler--time spectrograms. As $\sigma$ increases, the background texture becomes stronger, but the dominant UAV return and fall Doppler--time trajectory remain visible. This visual observation is consistent with the small accuracy changes reported in Table~\ref{tab:noise}.

\section{DISCUSSION}

The clean and noisy results point to a task-dependent quantum-kernel effect rather than a general performance claim. At $d=2$, the PCA bottleneck is severe and the RBF kernel is stronger on both tasks. Once the bottleneck expands to $d \in \{4,6,8\}$, the QSVC consistently improves the UAV benchmark, suggesting that the UAV RDM classes retain a geometry that benefits from the selected quantum fidelity-kernel mapping. The fall benchmark does not show the same separation. Its QSVC gains at $d=6$ and $d=8$ are small, and the two kernels remain close relative to the repeated-seed variability.

This difference is plausible because the two radar products encode different physical structure. The UAV RDM is an instantaneous range--velocity representation of airframe and rotor scattering, while the fall spectrogram is a Doppler--time representation of human motion evolution. A kernel that improves one representation need not improve the other. The result therefore argues against presenting the QSVC as uniformly superior. The more defensible conclusion is that the quantum fidelity kernel provides measurable utility on one controlled DT-generated radar task under a compact feature bottleneck, while the second task remains close to parity.

The noise experiment also clarifies what type of robustness is being tested. We inject additive Gaussian noise into the raw test arrays, but we do not perturb the training set, refit the PCA basis, simulate hardware noise, or model radar-specific clutter. The small changes in accuracy indicate that the trained compact pipeline is stable to this particular input perturbation. They do not establish robustness to measured-domain variability, adversarial perturbations, sensor miscalibration, multipath-rich clutter, or quantum-hardware imperfections.

Several limitations define the scope of the benchmark. First, both datasets are fully synthetic. The UAV dataset is generated by a high-frequency electromagnetic DT of three airframes, and the fall dataset is generated by a $60~\mathrm{GHz}$ FMCW DT of an indoor environment.\cite{Sayed2024MicrowaveMagUAVDigitalTwins,Sayed2024DronesReview,Sayed2024SensorsLettersISARDigitalTwins,Sayed2024TMTTMIMOUAV,Elbadrawy2023MapCon,ratto2026radar} A measured-data evaluation, ideally supported by calibrated sim-to-real methodology,\cite{trinh2026physics} is required before deployment-relevant claims can be made. Second, all quantum kernel evaluations use noiseless classical statevector simulation in Qiskit.\cite{qiskit} Gate noise, finite-shot statistics, decoherence, queue constraints, and device topology are not represented. Third, the inputs are magnitude radar products rather than raw complex I/Q data, so the quantum feature map is not operating on the original radar phase. Fourth, the study is limited to $d \leq 8$, both to keep statevector-kernel computation tractable and to remain consistent with near-term qubit constraints. These limitations are not methodological failures; they define this work as an early-stage DT-based benchmark rather than a deployment or hardware demonstration.

\section{CONCLUSION}

We compared a classical RBF kernel and a fidelity-based quantum kernel on two DT-generated FMCW radar tasks: UAV classification from RDMs and fall detection from Doppler--time spectrograms. Under matched preprocessing and PCA bottlenecks, QSVC outperformed RBF--SVC on the UAV task from $d=4$ onward and reached $0.941$ accuracy at $d=8$, while the fall task remained close to parity with only small QSVC gains at $d \in \{6,8\}$. Additive Gaussian noise injected into the raw test inputs produced little accuracy change in either branch, and the UAV QSVC advantage was preserved across the tested noise levels. These results support DTs as controlled testbeds for compact radar-QML benchmarking, while measured-data validation, hardware execution, radar-specific perturbation models, and phase-aware or I/Q-valued features remain necessary future work.

\acknowledgments

The authors acknowledge support from NSERC, MITACS, and Synopsys, and from CMC Microsystems, manager of the FABrIC project funded by the Government of Canada. Computational resources for digital-twin generation were provided by the Wireless Sensors lab at the University of Waterloo.

\bibliographystyle{unsrt}
\bibliography{references}

\end{document}